\documentclass[aps,pre,longbibliography,notitlepage,floatfix,nofootinbib,rmp]{revtex4-1}

\usepackage{bm,amsmath,amssymb,graphicx,stmaryrd,float,subcaption,upgreek,sidecap,MnSymbol,mathtools,lipsum}
\usepackage{epstopdf}
\usepackage{physics}
\usepackage[abs]{overpic}
\usepackage[justification=raggedright]{caption}
\usepackage{enumitem}

\usepackage{hyperref}

\usepackage{xcolor}

\begin{document}
	\title{Avoiding localization instabilities in rotary pleating}
	\author{Tian Yu}
	\email{yut3@sustech.edu.cn}
	\affiliation{Department of Mechanics and Aerospace Engineering, Southern University of Science and Technology, Shenzhen, China, 518055}
	\author{J. A. Hanna}
	\email{jhanna@unr.edu}
	\affiliation{Department of Mechanical Engineering, University of Nevada, Reno, NV 89519}
	\date{\today}

\begin{abstract}
Rotary pleating is a widely used process for making filters out of nonwoven fabric sheets.  This involves indirect elastic-plastic bending of pre-weakened creases by continuously injecting material into an accordion-shaped pack.  This step can fail through a localization instability that creates a kink in a pleat facet instead of in the desired crease location.
In the present work, we consider the effects of geometric and material parameters on the rotary pleating process.  We formulate the process as a multi-point variable-arc-length boundary value problem for planar inextensible rods, with hinge connections.  Both the facets (rods) and creases (hinges) obey nonlinear moment-curvature or moment-angle constitutive laws.  Some unexpected aspects of the sleeve boundary condition at the point of material injection, common to many continuous sheet processes, are noted.
The process, modeled as quasistatic, features multiple equilibria which we explore by numerical continuation. 
The presence of, presumably stable, kinked equilibria is taken as a conservative sign of potential pleating failure.  Failure may also occur due to localization at the injection point.  We may thus obtain ``pleatability surfaces'' that separate the parameter space into regions where mechanical pleating will succeed or fail.  Successful pleating depends primarily on the distance between the injection point and the pleated pack.  Other factors, such as the crease stiffness and strength relative to that of the facets, 
also have an influence.  
Our approach can be adapted to study other pleating and forming processes, the deployment and collapse of folded structures, or multi-stability in compliant structures.
\end{abstract}

\maketitle

\section{Introduction}\label{se:pleatingIntroduction}

Instabilities involving deformation localization are a fascinating class of problems at the juncture of mechanics, geometry, and dynamics that are important to the successful forming of materials into industrial products \cite{SemiatinJonas84}.
The role of geometry is particularly apparent when thin sheet materials undergo forming processes with significant bending.
There is a large body of literature on the combined stretching and bending of metallic sheets for automotive and other applications \cite{hu2002mechanics}, but the relevant bending-dominated processes tend to be geometrically constrained, so that forming limits and localization are discussed with respect to thinning or shear banding \cite{MarciniakKuczynski79, TriantafyllidisSamanta86}.  In contrast, the topic of the present study is an instability in which the bending deformations themselves are highly localized, and can be described as a curvature blowup or runaway crease formation in either a desired or undesired location in a sheet.  Stretching deformations in the sheet are negligible.

Rotary pleating of filters is a common elastic-plastic bending process in which sheet material with pre-weakened creases is continuously injected into a flexible, accordion-shaped pack \cite{hutten2007handbook}.  
In the present work, we consider how geometric and material parameters may conspire to localize bending into an unwanted kink in a pleat facet, instead of in the pre-weakened crease location.  
While recovery (springback) and, sometimes, thermal setting behavior can be important components of a pleating process, we do not address such issues.  Furthermore, we model only the forward trajectory of the process, so that the distinction between elastic and plastic response does not become an issue.

Despite much empirical knowledge about pleating in industry, there appears to be little or no discussion of the mechanics of rotary or other pleating processes in the scientific literature.
Furthermore, the nonwoven fabrics formed by rotary pleating are soft sheet materials for which there is little knowledge about bending behavior at high curvatures.
The mechanics of crease formation and recovery, and particularly of networks of facets and creases, govern the dynamic behavior of folded deployable structures \cite{PapaPellegrino08, satou2014local, Dharmadasa20} and pleated fabrics \cite{Holdaway60, Patkar15}, 
as well as the assembly of packaging or other folded structures 
 \cite{HalladayUlm39}.
Naturally occurring pleated structures include insect wings and plant leaves \cite{newman1986approach,kobayashi1998geometry}. 

In the present work, we employ a quasistatic, plane-strain formulation featuring a multi-point variable-arc-length boundary value problem, recast into a standard two-point form.  Into this we input nonlinear moment-curvature and moment-angle constitutive laws for facets and creases, modeled as planar inextensible rods connected by hinges.
Boundary conditions corresponding to material ejection from a sleeve, and resistance from a flexible pleated pack, provide interesting complications for us to address.  Numerical continuation allows us to explore a landscape of structural equilibria, some of which correspond to kinked geometries representing pleating failure through bending localization.  Another constraint on the process is the possibility of plastic deformation due to excessive curvature at the point of ejection.
In the present work, we do not perform an exhaustive parameter sweep, but present representative behaviors observed during our exploration of a limited process-relevant parameter range.

The problem formulation and method of solution are presented in Section \ref{se:methods} and Appendix \ref{appse:BVP}.  Section \ref{se:rotarypleatingresults} features results in the form of bifurcation landscapes and pleatability surfaces delineating the combinations of processing and material parameters that should lead to successful pleat forming in a rotary process. We conclude with some discussion in Section \ref{se:pleatingdiscussion}.

\section{Models and methods}\label{se:methods}

\subsection{Rotary pleating model}\label{se:rotarypleatingmodel}

Due to the large aspect ratio of pleated sheets and localization imperfections, we construct a planar model of the rotary pleating process, assuming uniformity across the width of the sheet.
Although pleating can be performed at either low or high velocities, as a first step towards understanding the mechanics, we model the process as quasistatic, and thus consider only static equilibria of the pleated structure.  The materials are lightweight, so the effect of gravity is neglected.

\begin{figure}[h!]
	\centering
	\captionsetup[subfigure]{labelfont=normalfont,textfont=normalfont}
	\begin{subfigure}[t]{0.45\textwidth}
		\centering
		\includegraphics[height=2.25in]{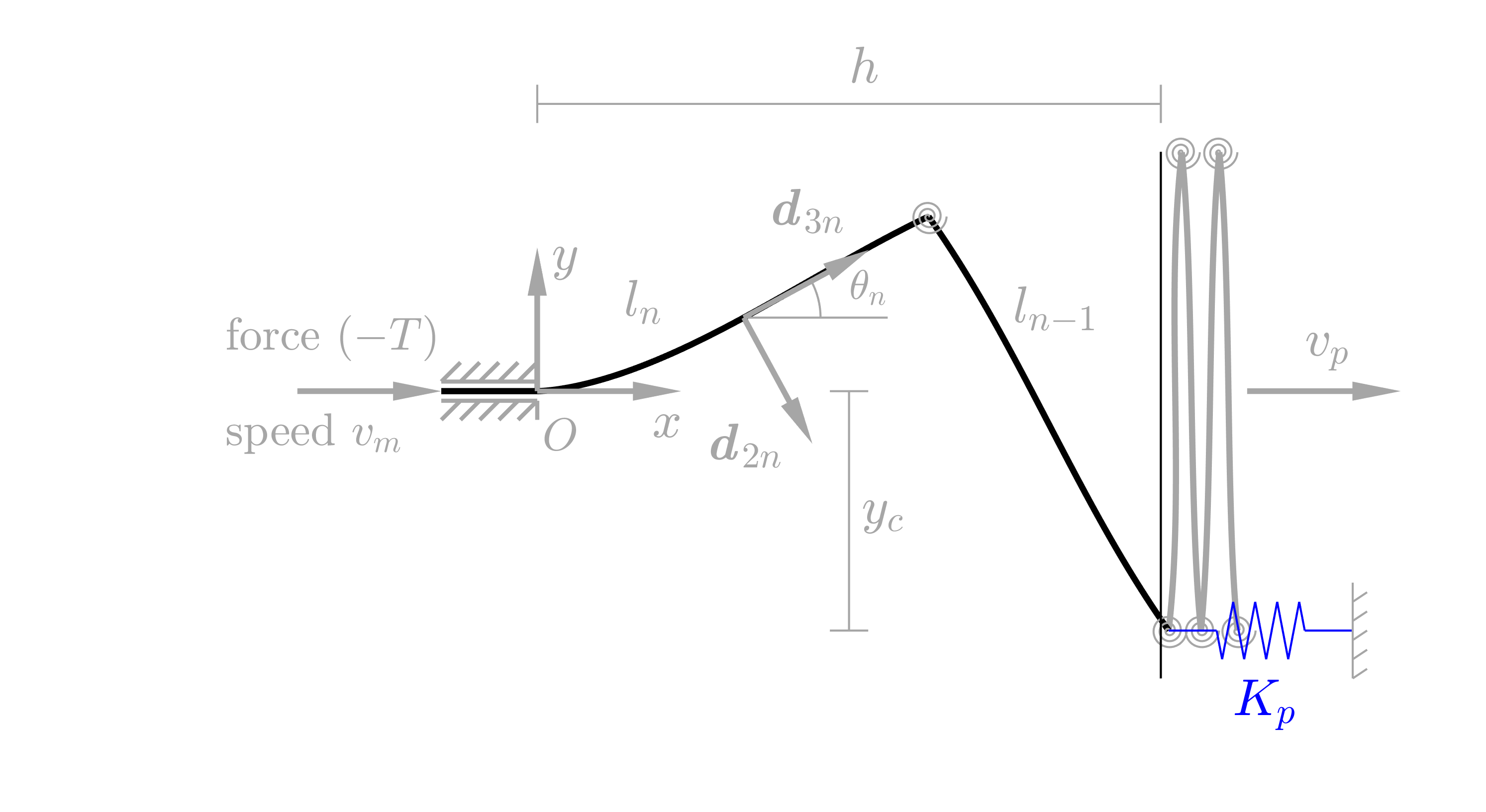}
		\caption{}\label{fig:rotarymodel}
	\end{subfigure}
		\hspace{.5in}
	\begin{subfigure}[t]{0.45\textwidth}
		\includegraphics[height=2in]{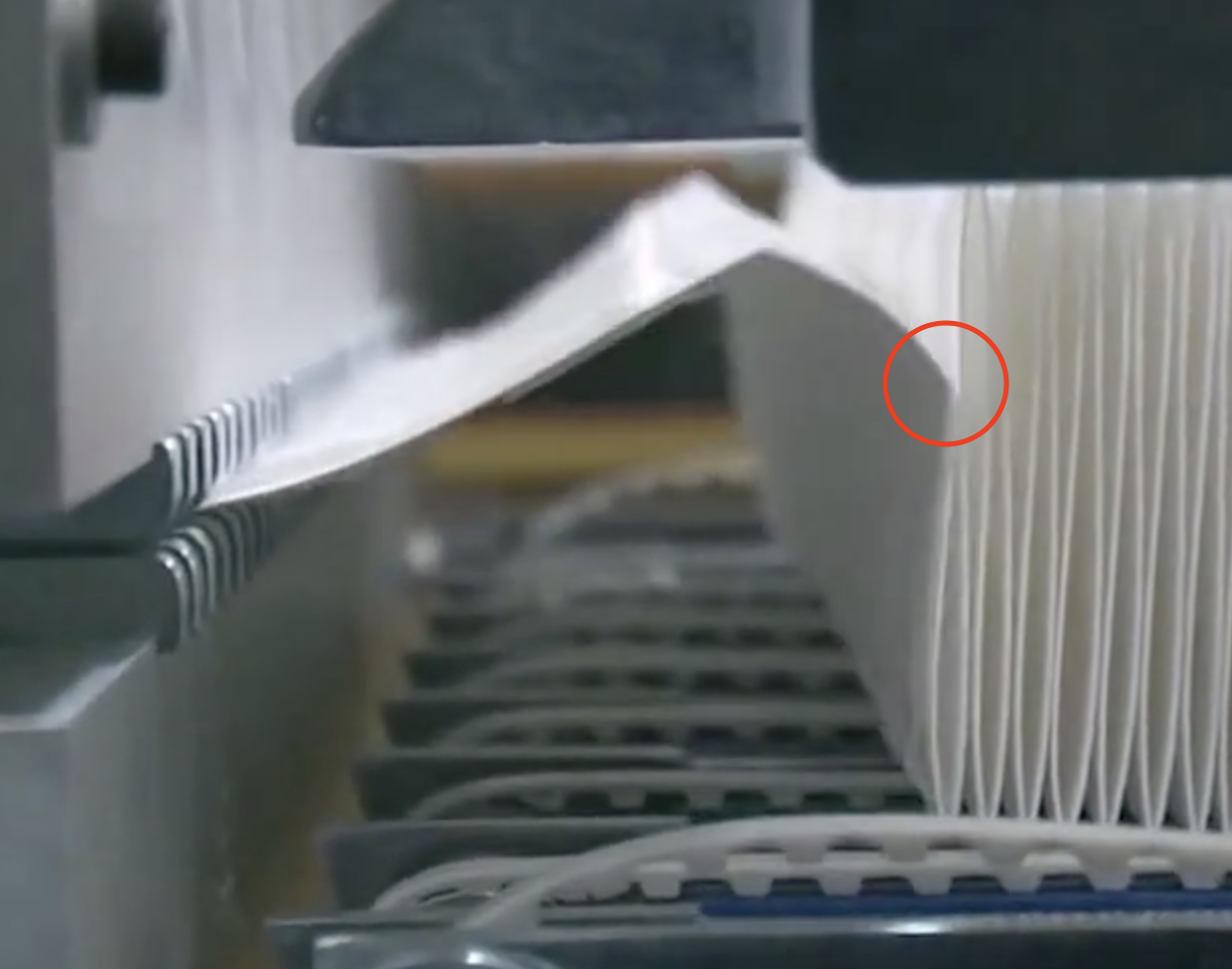}
		\caption{}\label{fig:rotarycal}
	\end{subfigure} 
	\caption{(\subref{fig:rotarymodel}) Rod and hinge model for rotary pleating, featuring multiple facets and hinges.  Material is ejected from the sleeve on the left, increasing the length of the leftmost segment $l_n$ from $0$ to $1$, after which a new crease and segment are inserted.  The material feels a (negative) tension $T$ at the sleeve edge.  Each facet is modeled as a rod with orientation $\theta$ and frame $\bm{d}_i$. Upon reaching a horizontal distance $h$ from the sleeve (vertical grey line), the outermost crease in the pack is set to a vertical distance $\approx y_c$ below or above the sleeve and coupled to a linear spring of stiffness $K_p$; other creases are moved rightward at a constant speed $v_p$ related to the final pitch of the pack.
 (\subref{fig:rotarycal}) A still from a video of a rotary pleating process \cite{rotarypleating}.  Inside the red circle is an example of transient curvature localization in an undesired location in the middle of a facet.}\label{fig:rotarypleating}
\end{figure}

Sheets are initially ``scored'' by rollers at predefined crease locations, after which they are ejected from a sleeve-like channel against a flexible, accordion-shaped pack of already-creased sheet sitting on a support \cite{hutten2007handbook}.  Ideally, this causes the sheet to localize curvature into sharp creases at the scored locations, in alternating directions.  However, sometimes the sheet will form a crease in the wrong place, such as in the middle of an un-scored facet, as shown in Figure \ref{fig:rotarycal}.

A schematic of our rotary pleating model is shown in Figure \ref{fig:rotarymodel}.  Pleat facets are separated into regular intervals by hinges, representing the pre-determined desired crease locations; several such segments of material are shown in the figure.  Each facet is modeled as a planar, inextensible rod with orientation $\theta$, curvature $\kappa$, and a moment-curvature constitutive law $M(\kappa)$. 
The creases supply one of the internal boundary conditions on the rod segments, and are themselves governed by a similar constitutive law for a hinge $M_c \left(\theta_{i-1}-\theta_i\right)$. 
We assign an index $i$ to each segment, with $i=n$ corresponding to the leftmost segment emerging from the sleeve.  This growing segment's length $l_n$ increases from $0$ to $1$, after which a new crease and segment are inserted.  All other segment lengths are unity.
Over to the right, the pleats enter the flexible pack.  In the actual process, facets contact each other in a complicated interaction in which the position of the boundary condition oscillates.  We mimic this in a simple way by adding a horizontal linear spring, of adjustable stiffness $K_p$ (normalized by facet stiffness and length), to resist the motion of the outermost crease in the pack once the crease has reached a horizontal distance $h$ from the sleeve.  
For the present paper
 we will report only results for $K_p \!=\! 80$, a dimensionless value chosen to mimic industrial settings by limiting the maximum displacement of the outermost crease to approximately the value of the normalized pitch of the final pleated pack.  
For multi-pleat calculations, we also move the other creases deeper in the pack rightward at a fixed speed $v_p$.   The ratio of the pack speed $v_p$ to the material injection speed $v_m$ is given by a geometric parameter related to the pitch of the final pleated pack; for the present study we set this ratio to $0.04$.  
All creases in the pack are set at a fixed height $\approx y_c = \pm0.5$ above or below the sleeve.  For multi-pleat calculations, this height is fixed by the initial vertical position of the crease when it reaches the horizontal distance $h$.  The process is thus very nearly symmetric with respect to the injection point.
Details of constitutive laws and boundary conditions will be presented in Sections \ref{se:constitutive} and \ref{se:rotarypleatingboundaryconditions}, respectively.

Each rod carries a right-handed orthonormal frame $(\bm{d}_1,\bm{d}_2,\bm{d}_3)$, with $\bm{d}_1$ pointing out of the plane and $\bm{d}_3$ along the tangent.
We need to solve for the components of the contact force $N_2$ and $N_3$ corresponding to the two in-plane frame directions. The contact moment $M(\kappa)$ around the out-of-plane direction is determined by the curvature $\kappa$.
The (negative) tension $T$ at the insertion point is an unknown.
The governing equations for the $i_{th}$ pleating segment are written in terms of derivatives with respect to a dimensionless length ${s}_i \in [0,1]$, the arc length of the segment normalized by $l_i$:

\begin{equation}\label{eq:2Drodequiscalar}
	\begin{aligned}
		N_{2i}' -N_{3i} \kappa_i l_i=0\,, \\
		N_{3i}' +N_{2i} \kappa_i l_i=0\,, \\
		M_i '(\kappa_i) - N_{2i} l_i =0 \,, \\
		\theta_i '=\kappa_i l_i \,, \\
		x_i '=l_i \cos \theta_i\,, \\
		y_i '=l_i \sin \theta_i \,.\\
	\end{aligned}
\end{equation}

This facilitates numerical solution of the equations by a continuation method.  This method and its initiation will be discussed in Section \ref{se:continuation}.

We find that the process reaches a steady state after insertion of only a few pleating segments.  In the end, we performed the parametric studies discussed in Section \ref{se:rotarypleatingresults} using a minimal model of only two segments, whose behavior is very close to the steady state behavior.  This saves time and avoids complications in the form of setting the height and speed of creases inside the pack.

\subsection{Constitutive laws}\label{se:constitutive}

The inextensible rod model of Equations \eqref{eq:2Drodequiscalar} can take as input any elastic moment-curvature relation $M(\kappa)$ for a sheet, as measured using a suitable bend tester \cite{Duncan99-1, Duncan99-2, yu2020exact}.  For this theoretical study, we use an adjustable function that mimics the observed elastic-plastic behavior of soft sheets during forward loading.  As we are not currently attempting to model any unloading or recovery of the sheet, we can treat the facets and hinges as effectively elastic, and need not track any state history of the material.

\begin{figure}[h!]
	\centering
	\includegraphics[width=0.9\textwidth]{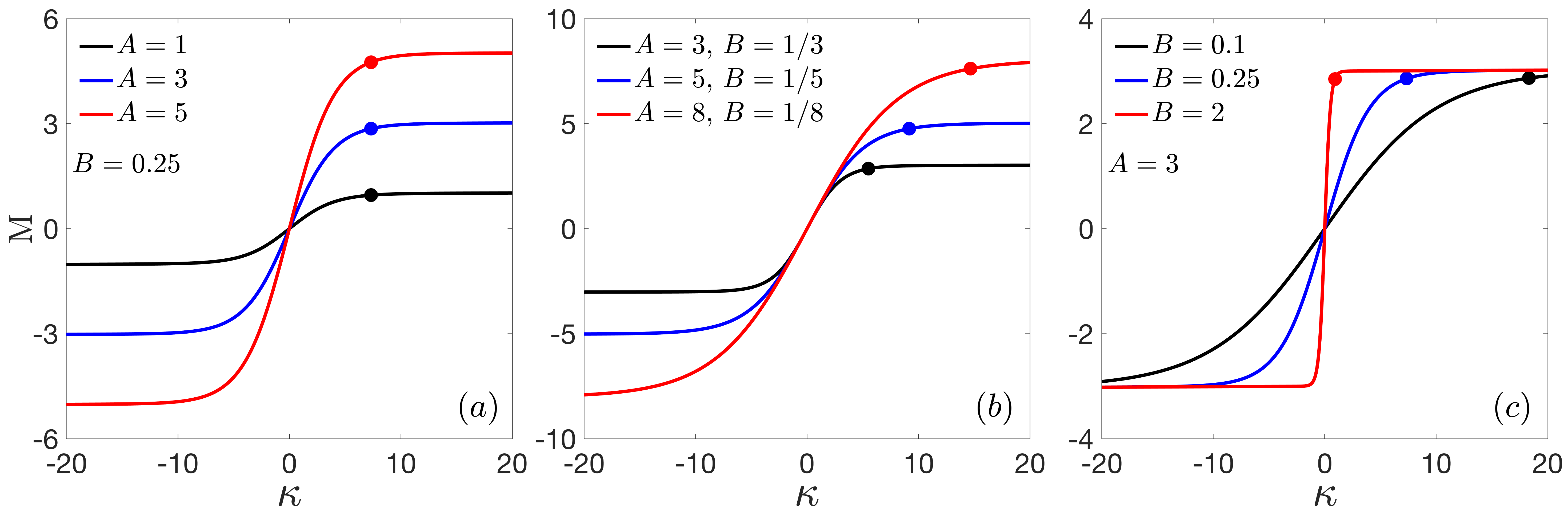}
	\caption{An adjustable facet constitutive law $M(\kappa )=A \tanh (B \kappa) +\epsilon \kappa$, with $\epsilon=0.001$.  Filled circles designate the assumed plastic localization curvature that occurs when $M =0.95A$. (a) Varying the plateau moment $A$ while fixing the localization curvature. (b) Varying the plateau moment and localization curvature while fixing the stiffness $AB$. (c) Varying the localization curvature while fixing the plateau moment.}
	\label{fig:tanh}
\end{figure}

The function we choose for the facets is a hyperbolic tangent, which mimics a material that approaches a plastic plateau stress after significant bending.
To stabilize the numerical continuation after curvature localization has initiated, we also add a very small linear term to this function so that it is always monotonically increasing, with a terminal slope of $\epsilon = 0.001$.  
Figure \ref{fig:tanh} shows how certain qualities of the function
\begin{align}
M( \kappa )=A \tanh(B \kappa) +\epsilon \kappa \label{constitutivelaw}
\end{align}
 can be adjusted, including its zero-curvature slope (stiffness) $AB$ and 
its plateau stress $A$.  Another important feature of a real soft sheet material is the curvature at which plastic deformation first occurs, which practical observations indicate is closely associated with the curvature at which localization begins.  For this study, we need an objective criterion to identify the onset of localization.  We choose the point at which the material has reached 95\% of its plateau stress, $0.95A$.  The condition $0.95A=A \tanh(B \kappa_{\mathrm{loc}})$ is equivalent to $\kappa_{\mathrm{loc}} \approx \pm 1.832/B$.
  This localization curvature can also be adjusted.

Similarly, we use the function
\begin{align}
	M_c\left(\theta_i , \theta_{i-1}\right) = A_c \tanh\left[B_c \left( \theta_{i-1}(0) - \theta_i(1) \right) \right] \label{constitutivelawcrease}
\end{align}
for the hinge response.
We thus define a nondimensional crease strength $C_\mathrm{stren} \equiv \tfrac{A_c}{A}$ and crease stiffness  $C_\mathrm{stiff} \equiv \tfrac{A_c B_c}{AB} \cdot 1$, where ``$1$'' is the unit length of a facet.  The length scale $\tfrac{AB}{A_c B_c}$ is akin to the ``origami length'' introduced in \cite{lechenault2014mechanical}, which describes the crossover between different regimes of behavior in creased sheets in which the predominant deformations are in the crease or in the rest of the sheet.
For the present study, we set $A=1$, so $B$ can be seen as a measure of facet stiffness although its true role is to control the localization curvature.  We consider only crease stiffnesses large enough that the crease moment reaches $0.95A_c$ before it bends from flat to a right angle; this corresponds to a range $\tfrac{C_\mathrm{stiff}}{C_\mathrm{stren}} B \geq 1.166$. We consider normalized 
crease strengths $C_{\text{stren}}$ up to 0.4. In industrial settings, scoring is achieved by compression of material, reducing its thickness locally.  Aside from any property changes due to this damage, the thickness reduction alone would account for a reduced bending stiffness which depends cubically on thickness, so that a crease with $40 \%$ of the strength of the original sheet would have been reduced in thickness by only about $25 \%$.

\subsection{Boundary conditions}\label{se:rotarypleatingboundaryconditions}

In our problem, there are conditions associated with the ejection of material from a sleeve as it enters the process window, internal conditions through which facets communicate through creases, and conditions that mimic the presence of a flexible pack at the end of the pleating process.  Within the pack, the conditions associated with the outermost crease differ from those deeper within the pack.

An interesting aspect of the boundary conditions associated with rods in frictionless sleeves is that the tension $N_3$ is not continuous across the sleeve edge \cite{bigoni15,oreilly2015eshelby, Hanna18}.  In the absence of plastic deformation or other dissipation, the quantity $N_{3}+M \kappa -\int M \kappa' ds$ will be conserved across the edge \cite{Hanna18}. 
With our constitutive law \eqref{constitutivelaw}, this corresponds to the following sleeve boundary condition:

\begin{equation}\label{eq:jumpcondition}
	T - N_{3n}(0)= A \kappa_n(0) \tanh(B \kappa_n(0))-\frac{A}{B} \ln [\cosh (B \kappa_n(0))] +\frac{\epsilon}{2} \kappa^2_n(0) \, .
\end{equation}
Additionally, we enforce continuity of position and angle $\theta$ at this boundary.

At any point in the process, there are one or two creases between the sleeve and the pack.  At these junctions, we enforce continuity of position, force (in $x$ and $y$ directions), and curvature.  The moment across the crease is also continuous, and equal to the crease moment $M_c$ given by the constitutive law \eqref{constitutivelawcrease}; this condition is enforced on only one side of the crease, otherwise the problem would be overconstrained.

Creases enter the pack when they reach a horizontal distance $h$ from the sleeve.   At this point, the vertical position of the crease is either fixed or set to the very close value of $-0.5$, this condition replacing force balance in the vertical direction. The outermost crease in the pack has its horizontal force balance augmented with a linear spring term of stiffness $K_p$ and rest position $h$. The other creases in the pack have their horizontal positions set to move at a fixed speed, in our case $0.04$ times the material injection speed, this condition replacing force balance in the horizontal direction. 
All other boundary conditions remain the same.  The leading edge of the first facet in the process is treated as having a crease with an angle corresponding to the final pitch, so that the function $M_c \left( \theta_1 , \arccos (0.04) \right)$ can be employed in the moment boundary condition.
The initiation of the process will be discussed in Section \ref{se:continuation}. 
The effect of a perfectly rigid pack can be achieved by treating the outermost crease in the same manner as the other creases in the pack.

\subsection{Continuation}\label{se:continuation}

The formulation presented in Sections \ref{se:rotarypleatingmodel}-\ref{se:rotarypleatingboundaryconditions} is a multi-point variable-arc-length boundary value problem.  Additional trivial ODEs are introduced in order to solve for unknown scalars appearing in the problem. 
  We employ the established technique of Ascher and Russell \cite{ascher1981reformulation} to recast this problem as a standard two-point boundary value problem, which we solve with the aid of the continuation package AUTO 07P \cite{doedel2007auto}. 
The technique assigns a full set of governing equations to every segment, normalizes the length of each segment to unity, and treats the matching or jump conditions between segments as boundary conditions. 
The method is illustrated, for the case of two segments, in Appendix \ref{appse:BVP}. 

For most of the calculation, we use the length of the emerging segment $l_n$ as a continuation parameter.  The unknown tension $T$ contributes an additional trivial ODE $T'=0$ to the system of equations \eqref{eq:2Drodequiscalar}.  When $l_n$ reaches unity, a new crease and (initially zero-length) segment are inserted at the sleeve edge.  The insertion of a crease involves an initially very large crease stiffness and a small constant added to the argument of the $\tanh$ within the crease constitutive law; 
 these values are continued to the correct crease stiffness and zero, respectively, before the injection of material is restarted.
The initiation of the pleating process is handled differently.  The starting configuration is horizontal, and the tension $T$ is set to zero.  While the leading edge is moved into its final vertical position in the pack, the unknown length of the emerging segment is solved for with a trivial ODE $l'_n=0$.  The final orientation of the leading edge is imposed after it reaches its final vertical position.

\section{Results}\label{se:rotarypleatingresults}

We consider the effects of several parameters on the pleating process, and here present observations of representative trends and behaviors. 
  The important parameters to be varied pertain both to the geometry of the problem as well as the material properties of the sheet.  The primary geometric parameter is the horizontal distance $h$ between the sleeve and pack, normalized by the unit facet length. 
Material parameters include the relative strength $C_\mathrm{stren}$ and stiffness $C_\mathrm{stiff}$ of the crease, and the plastic localization curvature $\kappa_{\mathrm{loc}}$ multiplied by the unit facet length.  Material and geometric effects are coupled through both the stiffness and localization curvature.
Another parameter, the stiffness of the pack $K_p$, was briefly explored but the results presented here are for a single, rather stiff value.
In our explorations, stiffer packs were found to behave similarly, with effects similar to a slight decrease in horizontal distance $h$.  Qualitatively different behaviors occur for much lower values of $K_p$ that are not industrially realistic; these were not systematically explored. 
Other geometric parameters are kept fixed in this study, including the pitch of the pack (equivalently, the speed of the pack) and the height of the creases with respect to the sleeve (equivalently, the up-down symmetry of the pack).
As mentioned previously, while we are able to model the formation of multiple pleat segments, we performed our parametric analysis on a minimal set of two segments and a single crease.  In this case, the pack pitch does not enter as a parameter at all. 

\begin{figure}[h!]
	\centering
	\includegraphics[width=0.7\textwidth]{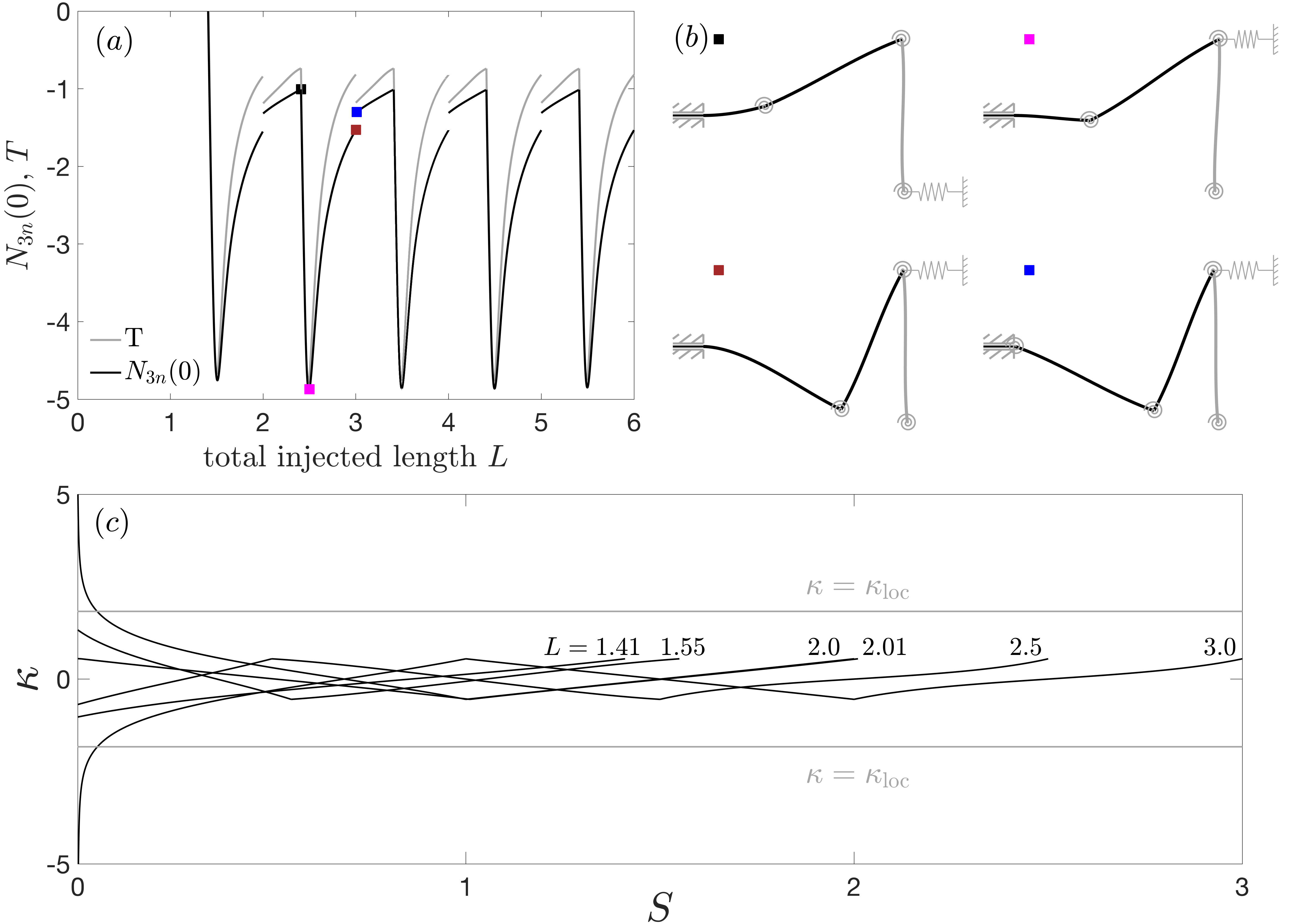}
	\caption{Rotary pleating of multiple pleats, showing the rapid approach to steady state behavior, using facet stiffness (inversely related to localization curvature) $B=1$, 
	 a horizontal spacing $h=1.3$, and a crease strength and stiffness $C_\mathrm{stren}=0.5$ and $C_\mathrm{stiff}=5$. 
	  (a) The compressive force $T$ inside the sleeve and the tension $N_{3n}$ immediately outside the sleeve.  Discontinuities correspond to crease insertion events at integer values of injected length $L$.  (b) Configurations corresponding to the markers in (a). (c) Curvature distribution along the arc length of the structure $0 \le S \le L$, measured from the sleeve edge,  for several values of total injected length $L$.
	 Failure occurs through the curvature $\kappa$ exceeding the localization value $\kappa_\mathrm{loc}$ near the sleeve edge; the off-scale values reach $14.4$ and $-15.8$.
	 Curvature relaxation due to crease insertion is apparent in both (b) and (c) (see the curvature relaxation at the sleeve edge from 2.0 to 2.01).}
	\label{fig:sleeveforce}
\end{figure}

\begin{figure}[h!]
	\centering
	\includegraphics[width=1\textwidth]{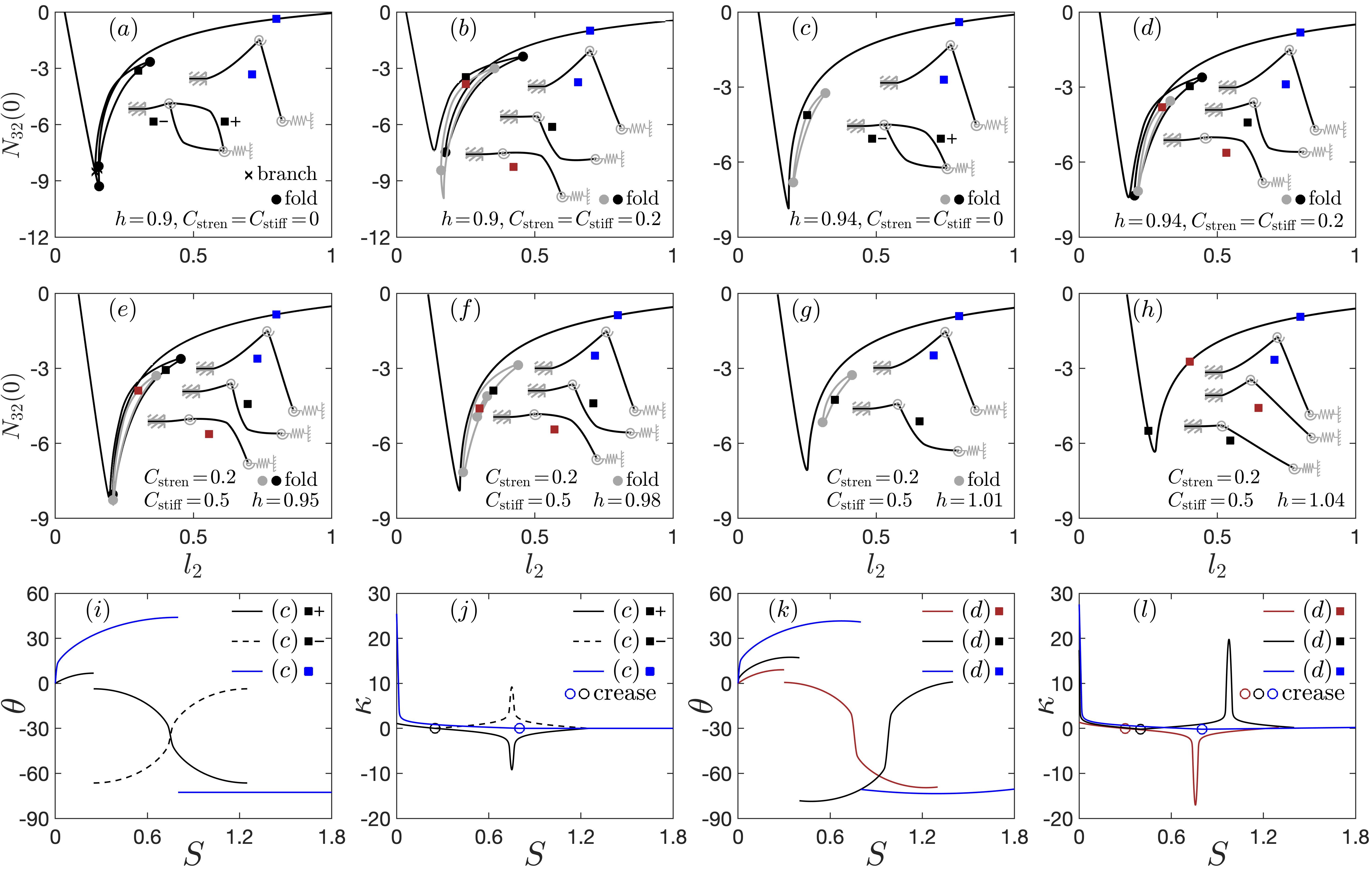}
	\caption{Results from a minimal rotary pleating model with facet stiffness (inversely related to localization curvature) $B=1$, 
	and different values of horizontal spacing $h$, crease strength $C_\mathrm{stren}$, and crease stiffness $C_\mathrm{stiff}$, as indicated in the subfigures. 
	 At a ``branch'' point, a pair of additional solutions bifurcate from the original branch.  
In (a)-(h) the bifurcation parameter is the length $l_2$ of the injected segment.  Physical configurations of the pleated structure are shown, corresponding to marked points on the equilibrium curves.  In addition to the main pleating branch, up to two connected (black) or disconnected (grey) looped branches appear that correspond to upward- or downward-kinked structures.
(i) Angle $\theta$ and (j) curvature $\kappa$ distribution of the insets in (c). 
(k) Angle $\theta$ and (l) curvature $\kappa$ distribution of the insets in (d). The horizontal axis in (i-l) measures the arc length along the structure from the sleeve edge.}
	\label{fig:bifurall}
\end{figure}

Results from a multi-pleat calculation are shown in Figure \ref{fig:sleeveforce}.  These include the tension inside and just outside the sleeve edge where material is ejected, configurations of the pleats, and the curvature distribution as a function of arc length $S$ along the structure. The process effectively reaches a steady state after only a few pleats are processed, with the behavior of the first pleat being quite close to that of subsequent pleats.  This justifies the use of our minimal model to assess the dependence of pleatability on various parameters. 
The compressive force $T$ inside the sleeve is always smaller in magnitude than the tension $N_{3n}$ immediately outside the sleeve \cite{liakou18constrained}. 
 Discontinuities in these values correspond to the insertion of new creases, which immediately relax the curvature $\kappa$ to a much lower value.  This example shows failure of the pleating process by a different mechanism than the localization occurring in Figure \ref{fig:rotarycal}, namely the attainment of unacceptably high curvature at the edge of the sleeve.

Both types of failure are present in several of the results shown in Figure \ref{fig:bifurall}, from a minimal model with facet stiffness $B=1$; this stiffness is inversely related to the localization curvature.
However, this figure is intended to illustrate certain transitions in the landscape of equilibrium configurations of the pleated structure that create or destroy states with kinked facets.
In Subfigures \ref{fig:bifurall}(a)-(h), the bifurcation parameter is the length $l_2$ of the injected segment and the response on the vertical axis is the (negative) tension just outside the edge of the sleeve.  Also shown are several physical configurations of the pleated structure corresponding to marked points on the bifurcation diagrams.   Note that as we are treating the material response as elastic, it only makes physical sense to move forward through any such diagram by increasing $l_2$, and only if the response is such that any local large values of curvature will monotonically increase.   Certain jump transitions that would occur at fold points in these diagrams do not meet the latter criterion.

In each diagram, there is a main branch of equilibria that continues up to the complete insertion of the segment.  Portions of this branch correspond to the progression of the desired pleating process, in which no curvature localization occurs in the middle of a pleat facet, with the caveat that pleating might still fail through the appearance of unacceptably high values of curvature near the injection point.  Up to two connected (black) or disconnected (grey) looped branches are present in Subfigures \ref{fig:bifurall}(a)-(g), and generally correspond to a pair of equilibria (one stable, one unstable) with localized curvature within a pleat facet.  Such kinks can be bent upward or downward; these do not behave symmetrically except in the limit of vanishing crease strength $C_\mathrm{stren}=0$.
The presence of such branches forms one conservative criterion for pleating failure.  If a kinked branch is connected to the main branch, the system will reach it smoothly in a quasistatic deformation.  But disconnected branches of equilibria can still be accessed by a real dynamic system possessing inertia and experiencing perturbations. 
The disappearances of the kinked solution loop branches occur through the collision and annihilation of two folds in the bifurcation landscape.  An efficient way to track this process is to examine the loci of these folds in parameter space.

In going from Subfigure \ref{fig:bifurall}(a) to \ref{fig:bifurall}(b), increasing the strength and stiffness of the crease disconnects one of two loops from the main branch, to form an isolated loop.
However, Subfigures \ref{fig:bifurall}(c) and \ref{fig:bifurall}(d) have a larger horizontal distance $h$ between the sleeve and pack, and in this case, increasing the strength and stiffness reconnects one of two loops.
In Subfigures \ref{fig:bifurall}(e)-(h), as the horizontal distance $h$ increases, loops disconnect and shrink, eventually disappearing.
Subfigures \ref{fig:bifurall}(i)-(j) and \ref{fig:bifurall}(k)-(l) show the distribution of angle $\theta$ and curvature $\kappa$ along the structure for several configurations from Subfigures \ref{fig:bifurall}(c) and \ref{fig:bifurall}(d), respectively.
At the hinges (open circles), the angle suffers a jump but the curvature is continuous.  At internal kinks, the angle changes rapidly and the curvature has a spike.  Note that in these examples, there is also a large curvature localized near the sleeve edge.

\begin{figure}[h!]
	\centering
	\includegraphics[width=0.85\textwidth]{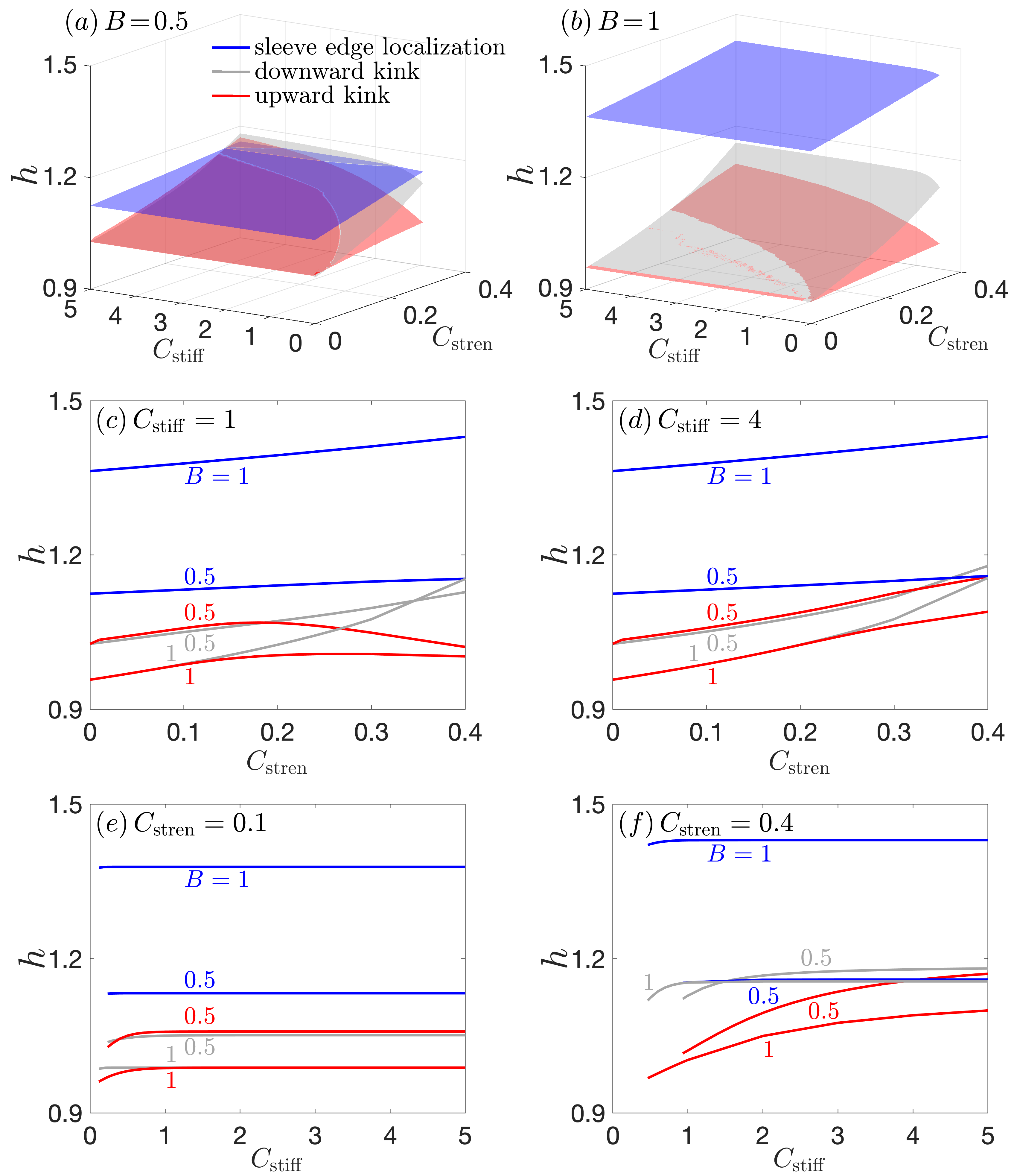}
	\caption{(a-b): Pleatability surfaces, showing $h$ as a function of crease strength $C_{\text{stren}}$ and crease stiffness $C_{\text{stiff}}$, corresponding to the elimination of kinked equilibria (red for upward kink and grey for downward kink) and avoidance of localization at the sleeve edge (blue) for 
	 two values of facet stiffness (inversely related to localization curvature) $B$.  As described in Section \ref{se:constitutive}, only values such that $\tfrac{C_\mathrm{stiff}}{C_\mathrm{stren}} B \geq 1.166$ are considered.
	 (c-f): Cross sections of the surfaces in two different $h-C_{\text{stren}}$ planes and $h-C_{\text{stiff}}$ planes.}
	\label{fig:pleatabilityAll}
\end{figure}

Figure \ref{fig:pleatabilityAll} presents, for two different values of $B$, three surfaces as functions of the crease strength $C_\mathrm{stren}$ and stiffness $C_\mathrm{stiff}$.  These surfaces represent conservative minimum values of horizontal spacing $h$ below which we expect pleating to fail by some mechanism.
Two pleatability surfaces correspond to the disappearance of two types of kinked loop branches (red for upward kink and grey for downward kink), and another simpler surface (blue) corresponds to failures from localization at the sleeve edge.  Illustrative cross sections of these figures are also shown. 
Recall that the facet stiffness $B$ is inversely related to the localization curvature.
 When $B=1$, pleatability is governed entirely by localization at the sleeve edge.  With $B=0.5$, the edge localization surface is much lower and, depending on the other parameter values, any of the three failure criteria can be the limiting factor for pleatability. In general, increasing the strength and stiffness of the crease makes kinking of a facet the more likely route to failure.
Starting at a given point in parameter space, failure can be avoided by either increasing the horizontal space $h$ or decreasing the pleat stiffness and the pleat strength. 
  As described in Section \ref{se:constitutive}, only the relevant part of the $C_\mathrm{stren}$-$C_\mathrm{stiff}$ plane was examined in this study.  The presence of a small ``lip'' in the red kink surface near the origin of this plane is noted, reflecting a small region of nonmonotonic dependence of the critical height on these parameters that we are not currently able to explain. 
Although not shown here, we found that rigidifying the pack slightly raises all the surfaces, particularly those associated with kinked equilibria, so that a greater horizontal spacing $h$ is required.

\begin{figure}[h!]
	\centering
	\includegraphics[width=0.85\textwidth]{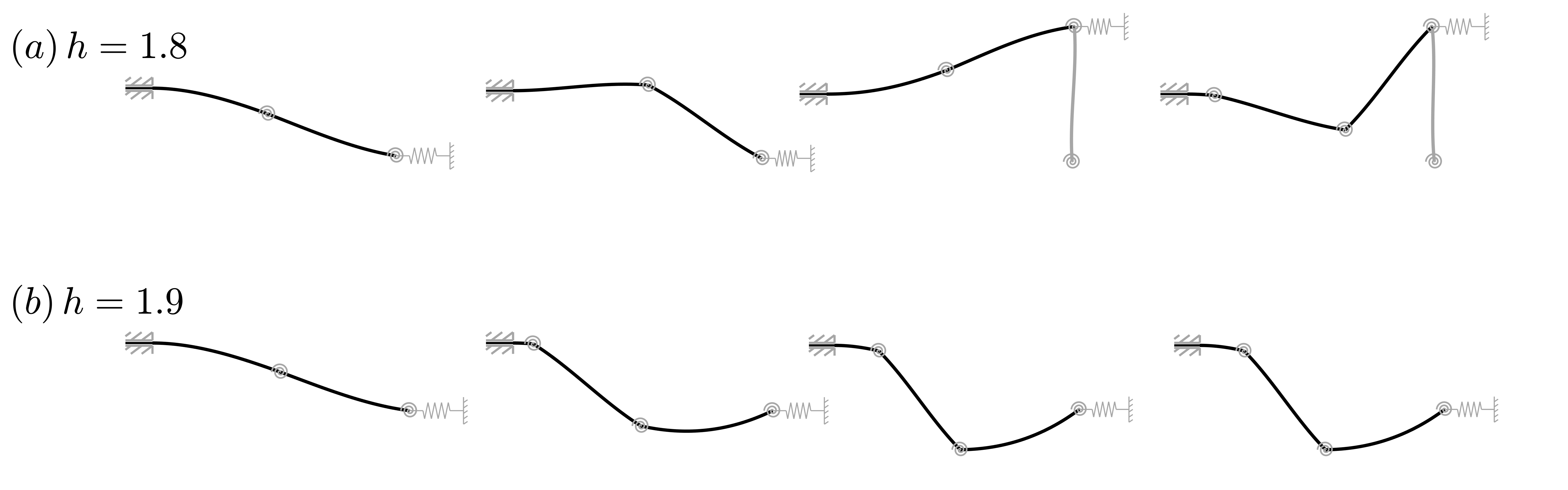}
	\caption{Pleating at large horizontal spacings. Parameters are facet stiffness (inversely related to localization curvature) $B=1$, 
	crease strength $C_\mathrm{stren}=0.5$, and crease stiffness $C_\mathrm{stiff}=5$.   
	(a) With a spacing $h=1.8$, pleating is successful. (b) With a spacing of $h=1.9$, a pleat misfolds in the wrong direction.  In a real pleating process, there would be a constraint below this configuration that would prevent the shapes shown here from fully forming.}
	\label{fig:Varylargeh}
\end{figure}

While increasing the horizontal spacing $h$ between sleeve and pack will simultaneously prevent different pleating failure mechanisms, the spacing cannot be increased indefinitely.  At too large a value of $h$, a crease may be injected before the previous one has deformed appreciably, and creasing may occur in the wrong direction.
The geometry of the problem is such that two full facets can fit as a straight line between sleeve and pack if $h = \sqrt{2^2-0.5^2} \approx 1.94$.  Continuation of equilibria suggest that creases will mis-bend at the slightly lower value $h \gtrsim 1.9$, as shown in Figure \ref{fig:Varylargeh}.

\section{Discussion and conclusions}\label{se:pleatingdiscussion}

Our results reveal a complex relationship between material properties and geometry of the structure and potential failure mechanisms during rotary pleating, although one that is in line with intuition.  
Greater facet stiffness or, equivalently, smaller localization curvature, tends to promote failure at the sleeve edge. 
Geometric confinement will induce sharper bending in the injected sheet, and it is then a matter of whether the weak point, the crease, yields so as to provide sufficient space before the maximum curvature in the pleat facet or at the sleeve edge exceeds the localization curvature.  It is interesting to note that for small enough horizontal spacing $h$, localization may occur even for vanishing crease strength and stiffness.  
A successful operation will relieve this geometric confinement as much as possible, within the constraints set by the requirement to isolate the deformation of each individual pleat.

Our criteria for avoiding facet and sleeve edge localization are both conservative.  It may not be necessary to completely eliminate the existence of disconnected localized solution loops in order to ensure that these kinked-facet solutions do not appear.  A more accurate criterion would require consideration of the role of inertia in accessing these or related states, and a careful stability analysis, incorporating the sleeve boundary condition, on the various solution branches.
With regard to sleeve edge localization, it is reasonable to assume that this effect will be relaxed if a more realistic imperfect sleeve with a finite gap is considered, a complication we did not attempt to explore.  Thus it is unclear how industrially relevant this type of localization is.  

Deeper study of these phenomena and validation of our assumptions will require controlled experiments with well-characterized materials. 
We have employed a simple phenomenological description of pleat facet and crease response. 
This elastic response cannot capture unloading behavior after localization, and some of the equilibria found through continuation may not be physically accessible.
Although we have treated the crease strength and stiffness as independent parameters, they are linked through the scoring process that weakens a pristine sheet to form the creases.
To obtain parameters related to the pleat facets, one can measure the moment-curvature response, up to the localization curvature, of a sheet using a bend tester \cite{Duncan99-1, Duncan99-2, yu2020exact}.  Measuring the response of the crease is, however, not a standardized process. 
Instead it seems that an analysis might be developed such that one or more simple loading processes could be used to determine the relevant nondimensional crease parameters.  However, one would still need to measure the localization curvature of the sheet material.

The implementation of our model can be adapted to study other pleating processes such as mechanical knife pleating \cite{hutten2007handbook}, which shares some features with a recently considered variable-length structural problem \cite{bosi2015self}, as well as other localization phenomena during sheet injection or growth processes. 
Finally, we note that sleeve boundary conditions are ubiquitous in continuous sheet processing, and so it is quite important to be aware of the, likely not well-known, presence of a tension jump at such injection points \cite{bigoni15,oreilly2015eshelby, Hanna18}.

\clearpage

\appendix

\section{Example standard form for two segments}\label{appse:BVP}

For illustrative purposes, we present the explicit formulation of the problem for the case of two segments, following \cite{ascher1981reformulation}. 
We have a multi-point variable-arc-length boundary value problem with additional unknown scalar quantities.  Continuation techniques require the standard form of a two-point boundary value problem over a unit interval. 

Both segments are normalized by their own lengths and are described by the same unit interval $[0,1]$. The length of the injected segment $l_2$ appears in the equations as a continuation parameter.  The length of the other segment $l_1 = 1$. 
Coupling at the internal pleat provides ``boundary conditions'' involving both boundaries, that is, the ``0'' end of segment 1 and the ``1'' end of segment 2.

We obtain a system of thirteen unknowns and boundary conditions.
The complete set of ODEs for the fields and scalars is

\begin{equation}\label{appeq:twosegs}
\begin{aligned}
&N_{21}' -N_{31} \kappa_1=0\,, N_{31}' +N_{21} \kappa_1=0\,, M_1 '(\kappa_1) - N_{21} =0 \,, 
 \theta_1 '=\kappa_1  \,, x_1 '= \cos \theta_1\,, y_1 '= \sin \theta_1 \,, \\
& N_{22}' -N_{32} \kappa_2 l_2=0\,, N_{32}' +N_{22} \kappa_2 l_2=0\,, M_2 '(\kappa_2) - N_{22} l_2 =0 \,,
 \theta_2 '=\kappa_2 l_2 \,, x_2 '=l_2 \cos \theta_2\,, y_2 '=l_2 \sin \theta_2 \,, \\
 &T'=0 \,, \\
\end{aligned}
\end{equation}
where the constitutive relations $M( \kappa_i )=A \tanh(B \kappa_i) +\epsilon \kappa_i $ ($i=1,2$) hold 
(compare equations \eqref{eq:2Drodequiscalar} and \eqref{constitutivelaw}).
The boundary conditions at the sleeve edge (see \eqref{eq:jumpcondition}) are
\begin{equation}\label{eq:BCrightmostpleat}
\begin{aligned}
&x_2(0)=0 , y_2(0)=0, \theta_2 (0) =0, 	\\
&T(0) - N_{32}(0)= A \kappa_2(0) \tanh(B \kappa_2(0))-\frac{A}{B} \ln [\cosh (B \kappa_2(0))] +\frac{\epsilon}{2} \kappa^2_2(0) \,. \\
\end{aligned}
\end{equation}
The boundary conditions representing the internal pleat (see \eqref{constitutivelawcrease}) are 
\begin{equation}\label{eq:ActivePleatsBC}
\begin{aligned}
& x_1(0)=x_{2}(1) \,, y_1(0)=y_{2}(1) \,,  \kappa_1(0)=\kappa_{2}(1) \,, \\
& A_c \tanh\left[ B_c \left( \theta_1(0) - \theta_2(1) \right) \right]  - A \tanh [ B \kappa_{2}(1) ] - \epsilon \kappa_{2}(1) =0 \,, \\
& N_{31} (0) \cos \theta_1(0) + N_{21} (0) \sin \theta_1(0) - N_{32} (1) \cos \theta_{2}(1) - N_{22} (1) \sin \theta_{2}(1) =0 \,, \\ 
& N_{31} (0) \sin \theta_1(0) - N_{21} (0) \cos \theta_1(0) - N_{32} (1) \sin \theta_{2}(1) + N_{22} (1) \cos_{2}(1) =0 \,.\\
\end{aligned}
\end{equation}
The boundary conditions at the pack (see discussion in Section \ref{se:rotarypleatingboundaryconditions}) are
\begin{equation}\label{eq:packoutermostpleat}
\begin{aligned}
& y_1(1)=-0.5 \,, \\
& A_c \tanh\left[ B_c \left( \theta_1(1) - \cos^{-1} (0.04)  \right) \right]  + A \tanh [ B \kappa_{1}(1) ] + \epsilon \kappa_{1}(1) =0 \,, \\
& K_p [(x_1 (1)-h] + N_{31} (1) \cos \theta_1(1) + N_{21} (1) \sin \theta_1(1) =0 \,. \\
\end{aligned}
\end{equation}


\bibliographystyle{unsrt}

\end{document}